\documentclass[review]{elsarticle}
\usepackage{lineno,hyperref,pifont,natbib,geometry,fleqn,graphicx,makeidx,fancyhdr,multirow,verbatim,amsmath,amsthm,amssymb,float,array,subfloat,epstopdf,subfig}

\journal{Journal of \LaTeX\ Templates}

\begin{document}

\begin{frontmatter}

\title{Characterization of Thin p-on-p Radiation Detectors with Active Edges} 


\author[a]{T. Peltola\corref{cor1}}
\ead{timo.peltola@helsinki.fi}
\author[b]{X. Wu} 
\ead{Xiaopeng.Wu@vtt.fi} 
\author[b,c]{J. Kalliopuska}
\author[d]{C. Granja}
\author[d]{J. Jakubek}
\author[d]{M. Jakubek}
\author[a]{J. H\"{a}rk\"{o}nen}
\author[a]{A. G\"{a}dda}

\address[a]{Helsinki Institute of Physics, P.O. Box 64 (Gustaf H\"{a}llstr\"{o}min katu 2) FI-00014 University of Helsinki, Finland}
\address[b]{VTT, Microsystems and Nanoelectronics, Tietotie 3, Espoo, P.O. Box 1000, FI-02044 VTT, Finland}
\address[c]{Advacam Oy, Tietotie 3, Espoo, FI-02150, Finland} 
\address[d]{Institute of Experimental and Applied Physics, Czech Technical University in Prague (IEAP-CTU), Horsk\'{a} 3a/22, CZ 12800 Prague 2, Czech Republic} 



\cortext[cor1]{Corresponding author}




\begin{abstract}
\label{Abstract}
Active edge p-on-p silicon pixel detectors with thickness of 100 $\mu\textrm{m}$ were fabricated on 150 mm float zone silicon wafers at VTT. By combining measured results and TCAD simulations, a detailed study of electric field distributions and charge collection performances as a function of applied voltage in a p-on-p detector was carried out. A comparison with the results of a more conventional active edge p-on-n pixel sensor is presented. The results from 3D spatial mapping show that at pixel-to-edge distances less than 100 $\mu\textrm{m}$ the sensitive volume is extended to the physical edge of the detector when the applied voltage is above full depletion. The results from a spectroscopic measurement demonstrate a good functionality of the edge pixels. The interpixel isolation above full depletion and the breakdown voltage were found to be equal to the p-on-n sensor while lower charge collection was observed in the p-on-p pixel sensor below 80 V. Simulations indicated this to be partly a result of a more favourable weighting field in the p-on-n sensor and partly of lower hole lifetimes in the p-bulk.  
%
%
%


\end{abstract}


\end{frontmatter}


\section{Introduction}
\label{Introduction}
An advantage offered by thin silicon radiation detectors in spectroscopic applications is the good radiation differentiation which is particularly important for nuclear industry where short range particles need to be detected with high gamma ray background \cite{Foulon99}. Also radiation hardness is important in many nuclear safeguard applications. Further benefits of thin detectors include reduced mass, fast charge collection and due to lower drift time for a given voltage, some possible advantage in charge collection and reverse current after high irradiation fluences \cite{Casse2009Cz}.  

The planar p-type silicon radiation detectors have been under extensive studies in the high energy physics (HEP) community due to their radiation tolerance and cost effectiveness, 
and have become a strong candidate to replace the conventional n-type detectors for the Large Hadron Collider (LHC) upgrade at CERN \cite{Casse2002p,Casse2006p}. 
A p-type detector is typically referred as n-on-p ($\textrm{n}^{+}/\textrm{p}^{-}/\textrm{p}^{+}$). 
The advantages of this configuration include a favourable combination of weighting and electric fields after irradiation due to the absence of type inversion. The readout at n-type electrodes enables the collection of electrons that have three times higher mobility and longer trapping times than holes, resulting in high speed readout and higher radiation hardness. Further asset of the p-type sensor is the reduced dependence of the charge collection efficiency (CCE) from the reverse annealing of the effective space charge in highly irradiated detectors \cite{Casse2006p,Kramberger2002p}.
%
%

However, the challenge of n-on-p design lies in the deteriorated isolation between the collection electrodes due to the electron accumulation layer induced by the positive charges in the $\textrm{SiO}_2$ passivation layer. Two methods, known as p-stop and p-spray, are widely used to improve the electrode isolation. Both methods, however, increase the process complexity and might lead to localized high electric fields which increase the likelihood of early breakdowns.

Thin p-on-p pixel detector addresses this problem without compromising CCE and spatial resolution excessively. Compared with the n-on-p detector, the p-on-p detector collects holes and the pn-junction remains on the unsegmented side of the detector. The relatively low hole mobility and long drift distance to the collection electrodes leads to a deteriorated CCE and make the thick p-on-p detector difficult to use. 
 
%
The motivation to implement the thin p-on-p pixel detector, except for the advantages mentioned above, is the expected good CCE after irradiation and an improved spatial resolution which make the thin p-on-p pixel detector distinguished from its thicker counterparts. The sensitive volume of the sensor can be further extended by using active edge or edgeless design that minimizes the regions where the signal cannot be collected \cite{Li2002,Kenney2006,Povoli2011,Macchiolo2014}.

This concept could be usable also for high energy physics (HEP) tracking applications. 
For instance, to maintain low material budget and achieve high position resolution the implementation of thin pixel detectors in the inner detector layers of the LHC experiments is foreseen for the future upgrades \cite{Terzo2014}. The thin active edge p-on-p configuration has not been studied before in the HEP community.

%
%
%
\section{Device and fabrication}
\label{Device and fabrication}
The investigated active-edge sensors were fabricated at VTT Technical Research Centre of Finland on 150 $\textrm{mm}$ Float zone silicon wafers from Topsil Semiconductor Materials A/S with a thickness of 100 $\mu\textrm{m}$. The geometry of the sensor is a matrix of 256 x 256 pixels with a 55 $\mu\textrm{m}$ pitch and a pixel implant diameter of 30 $\mu\textrm{m}$. The layer thicknesses and implantation depths as well as a cross-sectional view of the pixel implant and its metallization are presented in table~\ref{tab:2} and figure~\ref{fig:8c}, respectively. Both n-type (resistivity of 5 k$\Omega\cdot\textrm{cm}$) and p-type (resistivity of 10 k$\Omega\cdot\textrm{cm}$) wafers were used to fabricate sensors with different polarities. The basic electrical characterization ($IV$, $CV$, etc.) of both p-on-p and p-on-n sensors has been reported in the reference \cite{Kalliop2013}. The electrical contacts of the sensors are DC-coupled.
\section{Spectroscopic results}
\label{Spectroscopic results}

The p-on-p edgeless sensor was hybridized on the Timepix readout chip \cite{Llopart2007} at VTT 
with the lead-tin solder at a temperature of 210$^{\circ}$ C. The assembly was then wire bonded on the stacked PCB board designed by the Institute of Experimental and Applied Physics (IEAP-CTU), Czech Technical University in Prague. The FITPix \cite{Kraus2011} USB and Pixelman \cite{Turecek2011} software were used for the data acquisition and analysis. The detector was operated in the ToT (Time-over-Threshold) mode and with the bias voltage of 100 V.

%
In the Timepix detector, 
each pixel is connected to its individual preamplifier, discriminator and digital counter integrated on the readout chip. The detector works in one of three modes: Medipix mode (the counter counts incoming particles), Timepix mode (the counter works as a timer and measures the time when the particle is detected) and 
the ToT mode (the counter is used as a Wilkinson type ADC allowing direct energy measurement in each pixel). The Timepix detector working in ToT mode measures the charge collected from each pixel. As the device contains 65536 independent pixels and their response can never be identical, it is necessary to perform an energy calibration for each of them.

The energy calibration and the spectroscopic characterization were performed with the radioactive source $^{241}$Am and X-ray fluorescence emitted  by a mini X-ray tube with various target materials (Cu, Zn, Zr, Mo, Cd, In).

The p-on-p detector was first energy-calibrated with seven monoenergetic radiations. Only single pixel clusters were recorded to avoid the pile-up of signals. 
If the effect of charge sharing with the neighbouring pixels is sizeable, this will generate clusters including more than one pixel. For our investigation, all clusters comprising of more than one pixel were excluded. 
For the 100 $\mu\textrm{m}$ sensor with the bias above $V_\textrm{fd}$, the charge sharing is not a sizeable effect. 

\begin{figure}[tbp] 
\centering
\includegraphics[width=.45\textwidth]{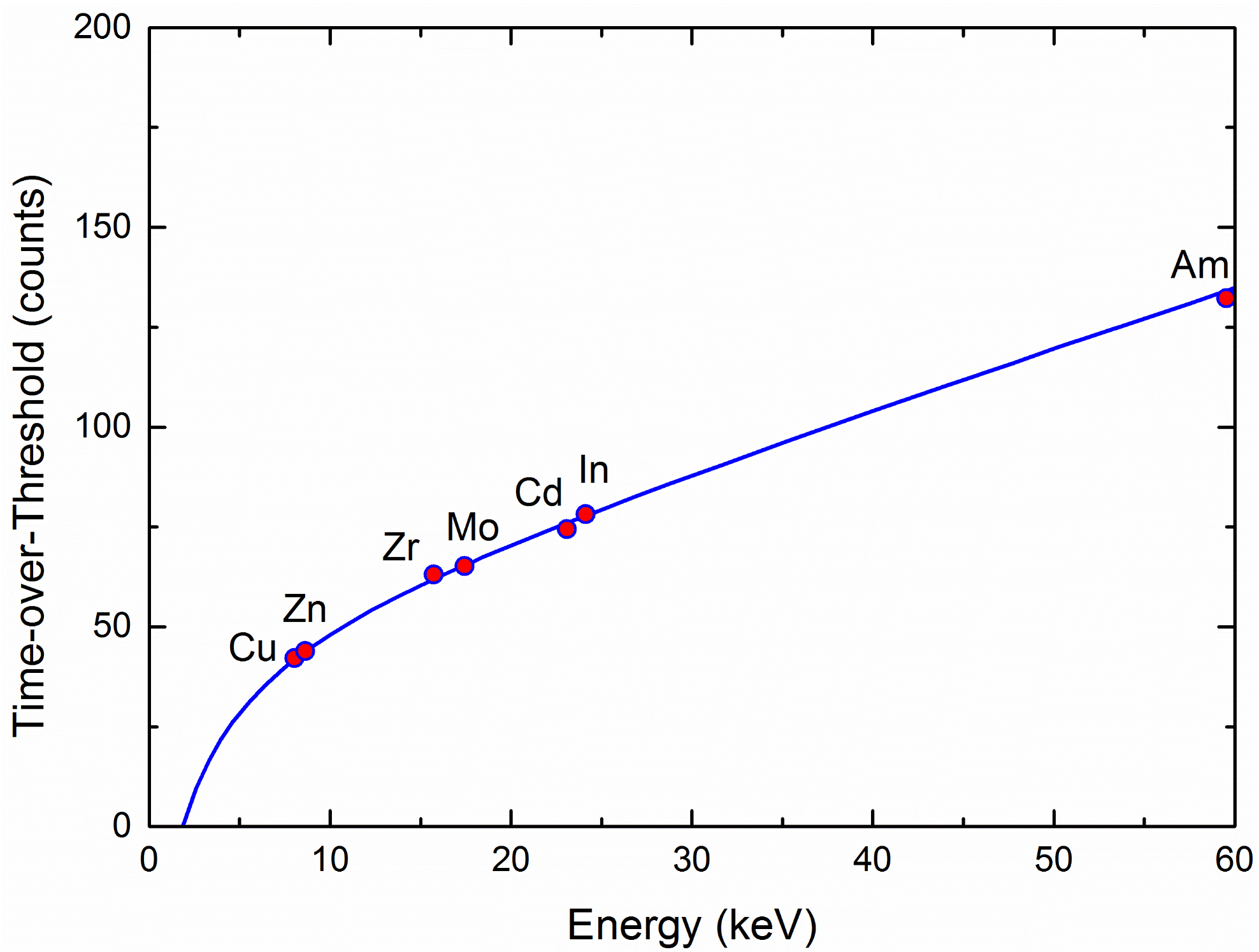}
\caption{Global energy calibration of the p-on-p edgeless detector hybridized on the Timepix readout chip.} 
\label{fig:fig1}
\end{figure}

The spectral peak positions were found by Gaussian fitting method. Then the calibration was performed by fitting the seven ToT peaks to the known energies, as reported in \cite{Jakubek11}. Figure~\ref{fig:fig1} shows the global energy calibration curve of the investigated p-on-p detector. In reality, the calibration process was performed for each pixel individually which results in 65536 calibration curves. The energy calibration was performed for both p-on-n and p-on-p detectors.

The edgeless technology used in the fabrication minimizes the inactive regions at the edges of the detector. The vicinity of n$^+$ doped edge, however, distorts the electric field distribution and thus might influence the CCE of the nearest pixels. 
To study the edge pixel dependence on the pixel to physical edge distance, the four detector borders were designed to have various pixel-to-edge distances (50 $\mu\textrm{m}$, 100 $\mu\textrm{m}$, 150 $\mu\textrm{m}$ and 200 $\mu\textrm{m}$).

\begin{figure}[tbp]
     \centering
     \subfloat[]{\includegraphics[trim=0.0cm 0.3cm 0.0cm 0.0cm, clip=true, width=.4\textwidth]{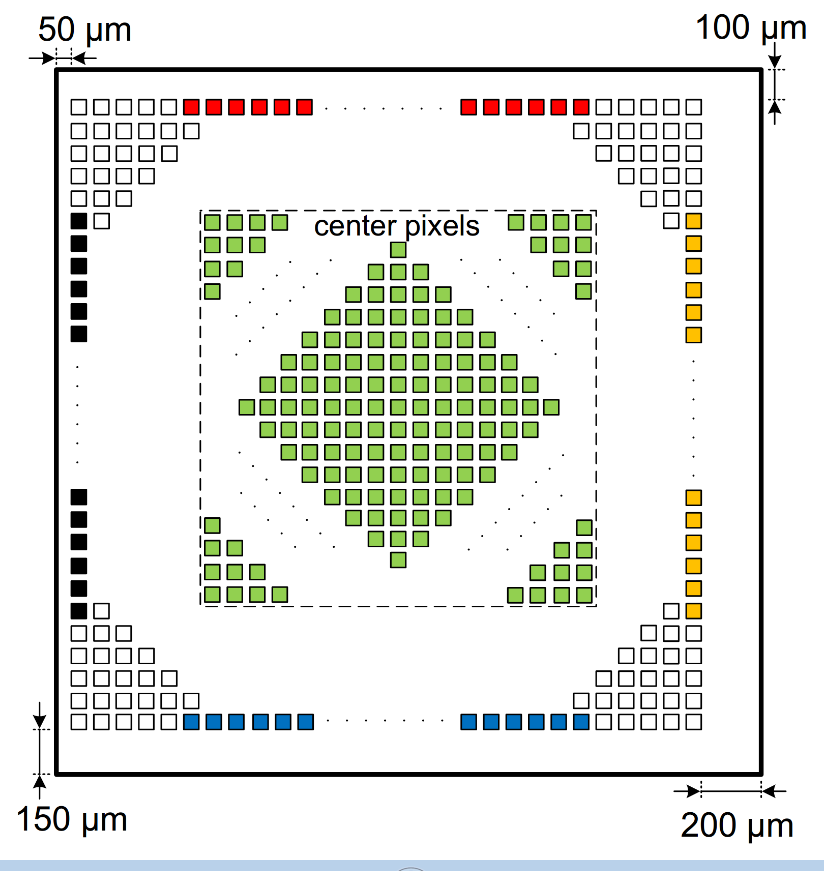}\label{pixel}}
     \subfloat[]{\includegraphics[width=.55\textwidth]{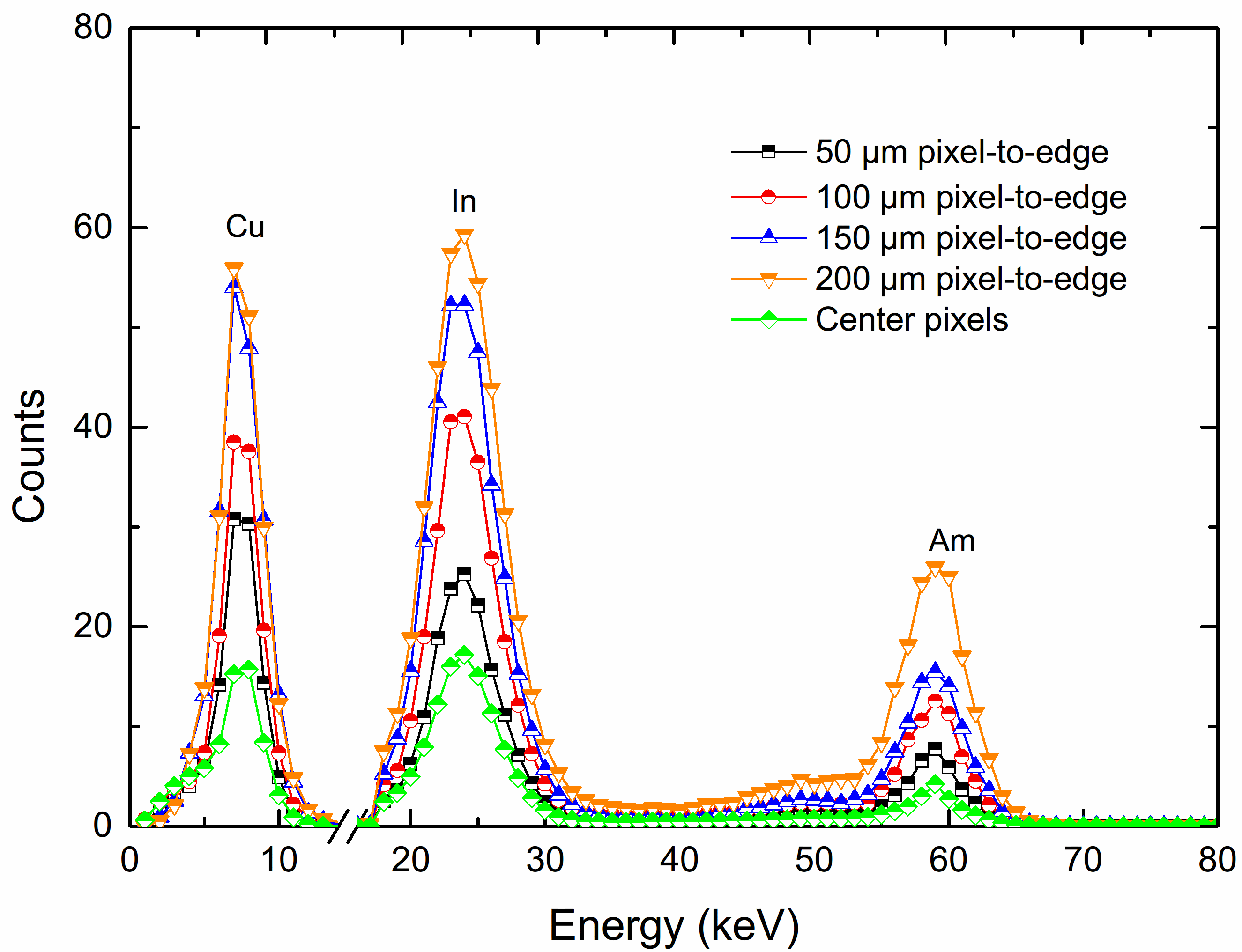}\label{spectrum}}
     \caption{(a) The $256\times256$ pixel topology of the detector (not to scale). The filled-in pixels were used for the response analysis of the given edge. 
The pixels in the center highlight the position of the pixels used to calculate the center response. For clarity a part of the pixels are not pictured, i.e. all the 
blank regions contain pixels and all the pixels within the central square were included in the center response. 
(b) Spectral response of the p-on-p edgeless detector to the copper and indium fluorescence
and americium irradiation. Only the mean values of single pixel events were studied.}   
     \label{fig:fig2}
\end{figure}

Figure~\ref{fig:fig2} shows the spectral responses of the edge pixels compared to the center ones. The spectral responses of the outermost pixels were summed along the edge and compared with the response of the center pixels. The responses of the corner pixel and its four neighboring pixels were excluded from the data analysis to eliminate the corner effect due to the wider depletion volume of the corner pixel and the 
electric field distortion due to the two edges. Since the center pixels have a smaller charge collection region this results in a smaller mean response than for the pixels at the edges. To reach sufficient statistics the data acquisition for e.g. Am exposure took about 30 minutes (168579 frames at 0.01 s per frame). Totally 20 million clusters were recorded resulting in $\sim$300 events for each pixel. As shown in the figure, the peaks of all the edge pixels and the center pixels are well aligned (sampling was done with the energy interval of 1 keV and all energy peaks appeared in the same location) and the heights of the peaks are proportional to the effective volumes of the pixels, indicating good functionality of the edge pixels. 
%

%
\section{Three-dimensional spatial mapping}
\label{Three-dimensional spatial mapping}

A 3D spatial mapping system at IEAP-CTU \cite{Jakubek2013_3D,Wu2014_3D} was used to investigate the influence of the active edge on the charge collection volumes of the edge pixels. The principle of the scanning system is to use a collimated tungsten X-ray (40 kV) penetrating the detector in a sharp angle (70$^{\circ}$), allowing the beam with a diameter of less than 1 mm to interact with several pixels at various depths. When the detector is shifted perpendicularly to the beam direction, the interactions at all pixel depths are recorded and a 3D map of the detector charge collection volume is obtained.
%

\begin{figure}[tbp] 
\centering
\includegraphics[width=.45\textwidth]{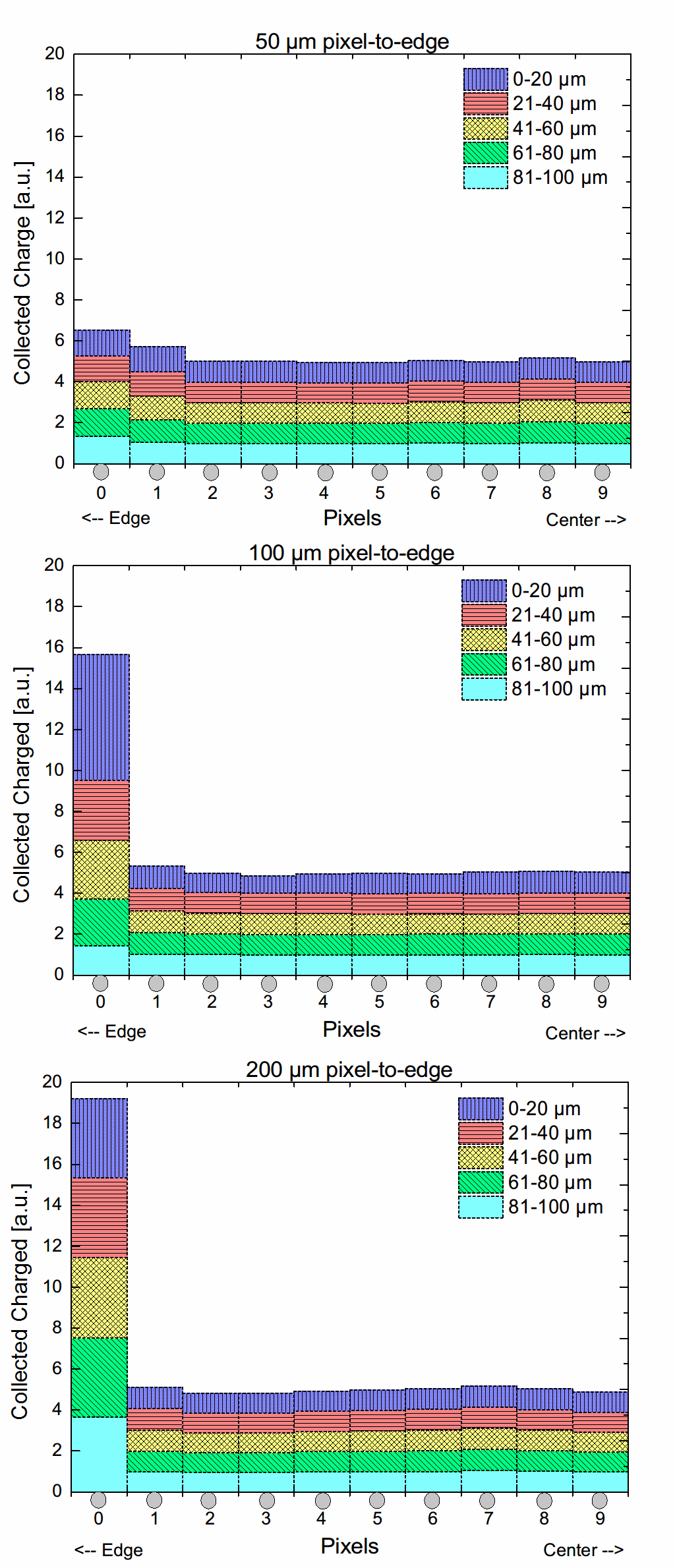}
\caption{Spatial mapping results of the edge pixels at three borders. The colors represent different depths with respect to the segmented detector surface. The outermost pixels in the diagram have 50, 100 and 200 $\mu\textrm{m}$ distance to the physical edge.} 
\label{fig:fig3}
\end{figure}

Figure~\ref{fig:fig3} shows the spatial mapping results of the p-on-p edgeless detector at a bias voltage of 20 V. The voltage was selected to be above the $V_{\textrm{fd}}\approx10$ V with some margins. Thus, the depletion volume will reach the backplane of the 100 $\mu\textrm{m}$ thick sensor but the lateral expansion will have different distances to the edge. 
Ten pixels adjacent to the detector edges were investigated. 
The X-ray enters the detector with a certain angle. Therefore the charges collected from different interaction depths of the pixel can be recorded. These regional signal responses are corresponding to the sensitive volumes of the pixels at certain depths which are indicated by different colors in figure~\ref{fig:fig3}. It can be seen that the outmost edge pixels collect charges from wider volumes than the other pixels. The pixels having wider pixel-to-edge distances received correspondingly more charges, indicating that the electrical distortion happened near the doped edges that extended the depletion volume towards the edges. 

When the measurement results from the three sensors in figure~\ref{fig:fig3} are compared, it can be observed that the amount of charge registered in the outmost edge pixel is not proportional to the physical edge width of the sensor. This is usually due to the existence of certain non-depleted volume in the sensor bulk, e.g. at the corner on the segmented side of the chip, where the electric field is too "weak" to build up 
a depleted region.
%

%

\section{Proton beam irradiation}
\label{Proton beam irradiation}
The proton beam tests were carried out at the Van de Graaff accelerator laboratory of the Institute of Experimental and Applied Physics\footnote{http://aladdin.utef.cvut.cz/projekty/VdG} in Prague. Accelerated protons of different energies from the beam line were used for the test. One 100 $\mu\textrm{m}$ thick p-on-n edgeless detector and one 100 $\mu\textrm{m}$ thick p-on-p edgeless detector were chosen for the test. The detectors and readout electronics were placed into a vacuum chamber into which the beam was guided. A gold foil (thickness in the range of 0.5 mm) was positioned in the vacuum chamber towards the beam direction to scatter the high intensity proton beam to the entire area of sensor, i.e. the tilted angle of the foil with respect to the beam allows the proton scattering to the sensor. The detectors were irradiated by the scattered beam from the non-segmented backplane and the signal was transmitted to the DAQ with coaxial cables. 

\begin{figure}[tbp] 
\centering
\includegraphics[width=.65\textwidth]{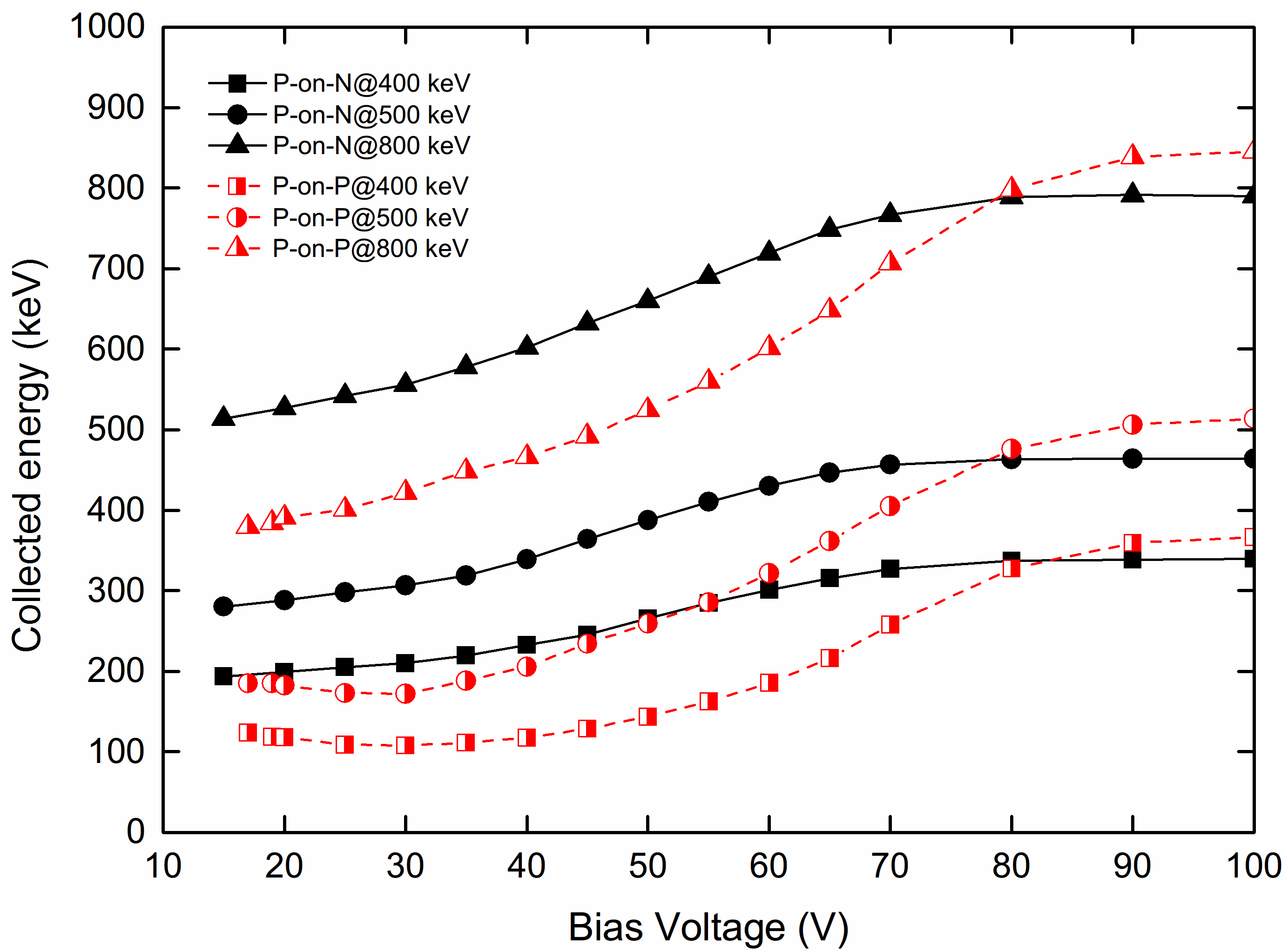}
\caption{
Collected proton energies (charges) as a function of the applied bias voltage.}  
\label{fig:fig4}
\end{figure}

\begin{figure}[tbp] 
\centering
\includegraphics[width=.65\textwidth]{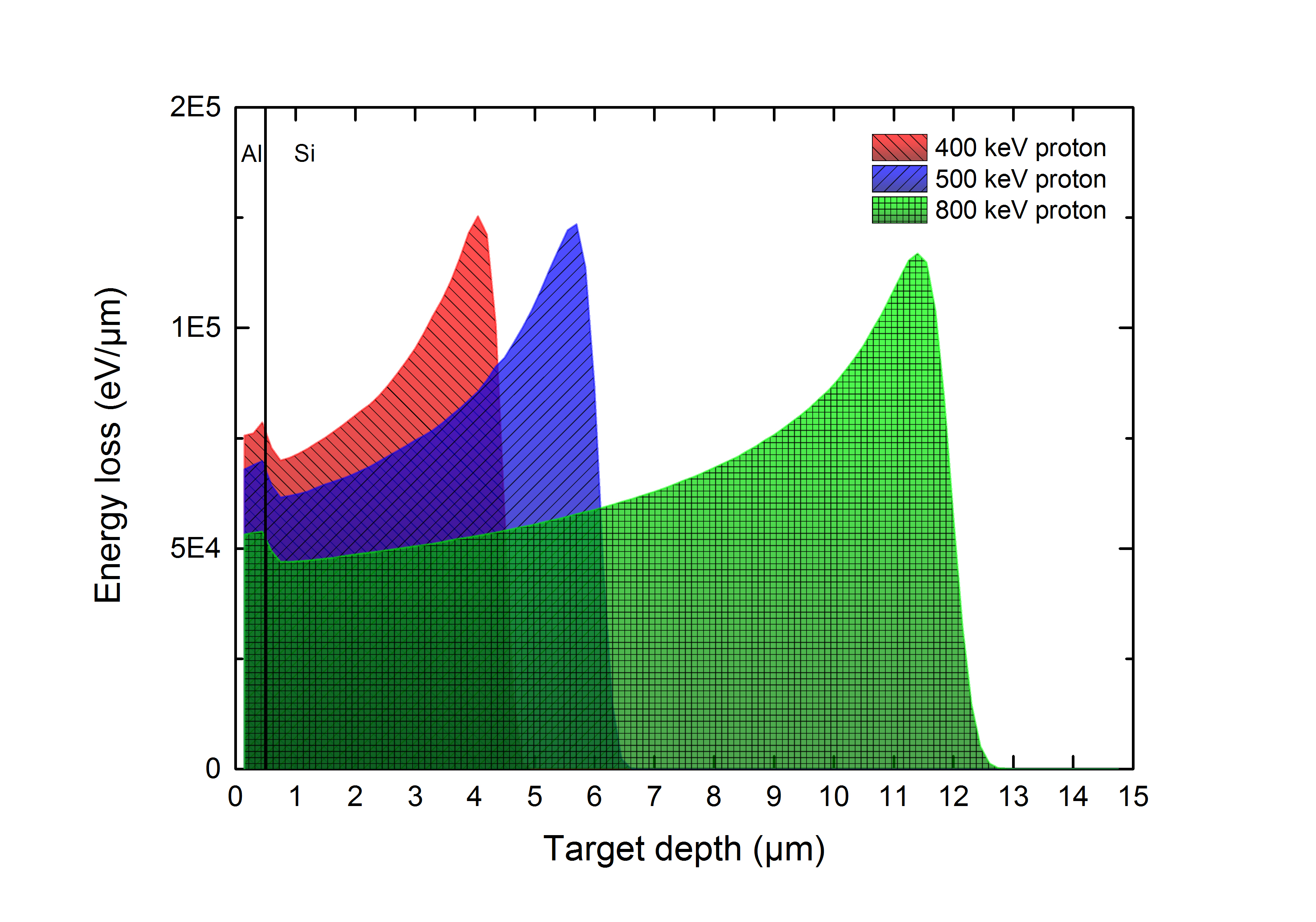}
\caption{The SRIM simulated absorption depths in silicon with a layer of aluminum as in the backplane of the measured detectors.} 
\label{fig:fig5a}
\end{figure}

Figure~\ref{fig:fig4} shows the deposited energies on two edgeless detectors 
as a function of applied bias voltage. It can be seen that the energies collected by the two 100 $\mu\textrm{m}$ thick detectors increase with the bias voltages and the correct energies are registered only when the bias voltage is above 80 V. Also the p-on-n detector is collecting more charge for the most part of the voltage range. This behaviour is further investigated in section~\ref{TCAD Simulations}. 
Figure~\ref{fig:fig5a} shows a simulation using the SRIM tools\footnote{www.srim.org}. The 300-800 keV protons are mostly absorbed within the 12 $\mu\textrm{m}$ depth from the incident silicon surface. 
%
%
%
%

%
\section{TCAD simulations and modelling}
\label{TCAD Simulations}
The simulations presented in this paper were carried out using the Synopsys Sentaurus\footnote{http://www.synopsys.com} finite-element Technology Computer-Aided Design (TCAD) software framework.
\subsection{Simulation set-up}
\label{Simulation set-up}
\begin{figure}[tbp] 
\centering
 \includegraphics[width=.7\textwidth]{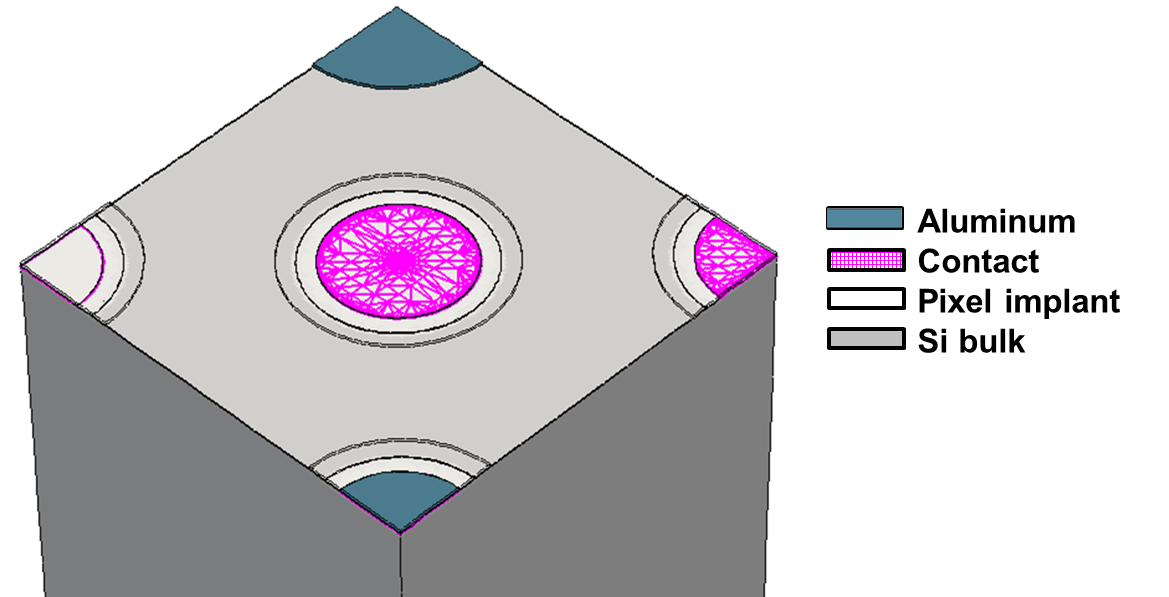}
\caption{
%
Simulated 3-dim. pixel sensor front surface structure of 5 pixels with oxide layer stripped, showing different details in each pixel. The mesh regions 
at the center and rightmost pixels highlight the positions of the DC-coupled contacts. The gray layers on the top and bottom corners are the pixel metallization and 
via structures, respectively. Black contour lines around the four pixels illustrate the edge of the stripped aluminum overhang. Pixel implants are shown in white 
while the lightly doped Si bulk is shown in light gray. The n$^+$ layer and the metallization of the non-segmented backplane are not pictured.}
\label{fig:8b}
\end{figure}

For the simulation study of the observed charge collection behaviour in figure~\ref{fig:fig4} the 3-dimensional structure presented in figure~\ref{fig:8b} was applied. This was deemed necessary since in a 2-dim. structure only interactions between a single column of pixels would be monitored. Also the correct reproduction of the local electric fields due to the circular shape of the pixels required a 3-dim. structure. Since most of the pixels in the detector do not experience any influence from the active edges, these were not included in the simulated structure, as can be seen from figure~\ref{fig:8b}.
%

\begin{figure}[tbp] 
 \centering
  \includegraphics[width=.8\textwidth]{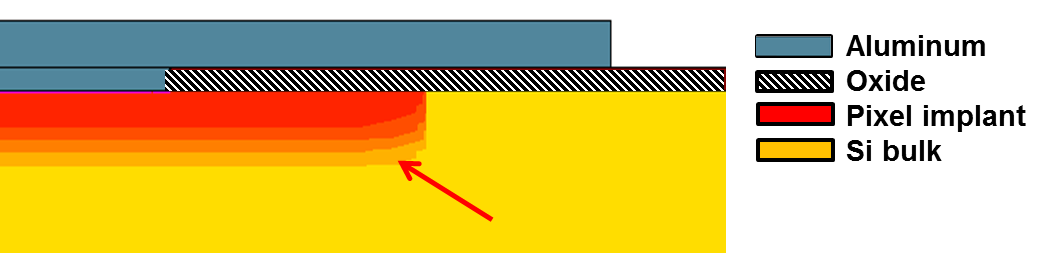}
 \caption{
Cross-sectional view of the simulated pixel implant and its metallization, showing from top to bottom: the aluminum layer, the oxide layer and the 
via-structure, and the heavily doped implant region (indicated by an arrow) surrounded by the lightly doped silicon.}
 \label{fig:8c}
\end{figure}

The simulated pixel sensor configurations, p-on-p and p-on-n, were designed with parameters as close to the real sensors as possible and the both sensor types had a physical thickness of 100 $\mu\textrm{m}$, a pitch of 55 $\mu\textrm{m}$ and a pixel implant diameter 
of 30 $\mu\textrm{m}$. The layer dimensions and doping parameters are given in table~\ref{tab:2}, in which the bulk dopings were estimated by using the resistivity data of the two sensor substrates ($\sim$10 k$\Omega\cdot\textrm{cm}$ for the p-on-p and $\sim$5 k$\Omega\cdot\textrm{cm}$ for the p-on-n). The lateral diffusion of the pixel implants was set to 0.8$\times$depth. The aluminum metallizations above the pixel implants and their vias through the oxide layer had the diameters of 36 $\mu\textrm{m}$ and 24 $\mu\textrm{m}$, respectively. Detailed cross-sectional slice of the pixel is presented in figure~\ref{fig:8c}. Since the only differences between the sensor types (both had p$^+$ implantations at the pixels and n$^+$ at the non-segmented side) were the pixel implant diffusion depths and the type and concentration of the bulk doping, the figures~\ref{fig:8b} and~\ref{fig:8c} can be considered to represent both of the simulated sensor structures. 

Each pixel had a DC-coupled electrode at zero potential with sufficiently low resistance for charge collection. The reverse bias voltage was provided by the backplane contact.
%
%
%
\begin{table}[tbp]
\caption{The layer dimensions and doping parameters for the two simulated sensor types. 
The $d_{\textrm{\tiny Ox,Al}}$ are the oxide and aluminum layer thicknesses, respectively. 
The depths and peak concentrations for the Gaussian decay of the heavily doped implantations to the bulk level are also given.}
\label{tab:2}
\smallskip
\centering
\begin{tabular}{|c|cccccc|}
    \hline
    {\small {\bf Sensor type}} & \multicolumn{1}{c|}{\small {\bf $d_{\textrm{\tiny Ox}}$}} & \multicolumn{1}{c|}{\small {\bf $d_{\textnormal{\tiny Al}}$}} & \multicolumn{1}{c|}{\small {\bf n$^{+}$ depth}} & \multicolumn{1}{c|}{\small {\bf p$^{+}$ depth}} & \multicolumn{1}{c|}{\small {\bf n$^{+}$/p$^{+}$ peak conc.}} & {\small {\bf Bulk conc.}}\\
     & \multicolumn{1}{c|}{\small \textnormal{[$\mu\textrm{m}$]}} & \multicolumn{1}{c|}{\small \textnormal{[$\mu\textrm{m}$]}} & \multicolumn{1}{c|}{\small \textnormal{[$\mu\textrm{m}$]}} & \multicolumn{1}{c|}{\small \textnormal{[$\mu\textrm{m}$]}} & \multicolumn{1}{c|}{\small \textnormal{[cm$^{-3}$]}} & {\small \textnormal{[cm$^{-3}$]}}\\
    \hline
    {\small p-on-p} & {\small 0.25} & {\small 0.5} & {\small 1.25} & {\small 0.75} & {\small $5\times10^{18}$} & {\small $1.326\times10^{12}$}\\
    \hline
    {\small p-on-n} & {\small 0.25} & {\small 0.5} & {\small 1.25} & {\small 0.80} & {\small $5\times10^{18}$} & {\small $0.883\times10^{12}$}\\
    \hline
\end{tabular}
\end{table}
%
%
%
%
\subsection{Simulation results}
\label{Simulation results}
\subsubsection{Electrical characteristics}
\label{Electrical characteristics}
\begin{figure}[tbp]
     \centering
     \subfloat[]{\includegraphics[width=.485\textwidth]{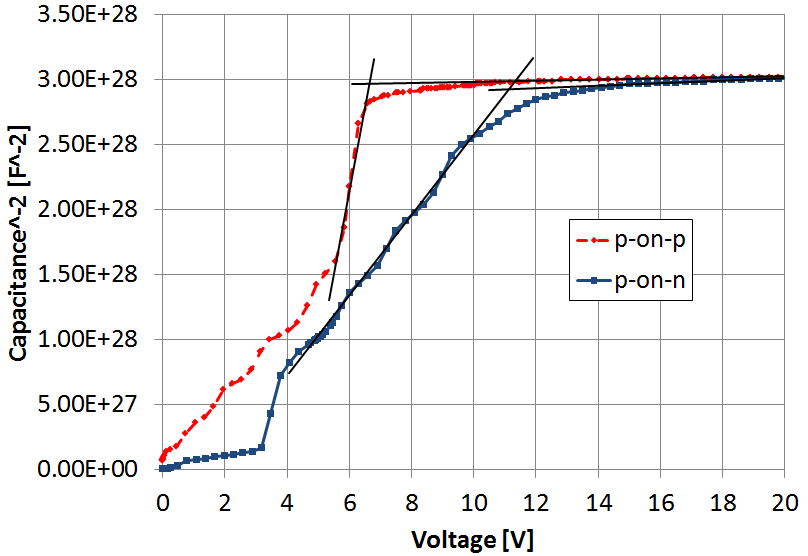}\label{CV}}
     \subfloat[]{\includegraphics[width=.47\textwidth]{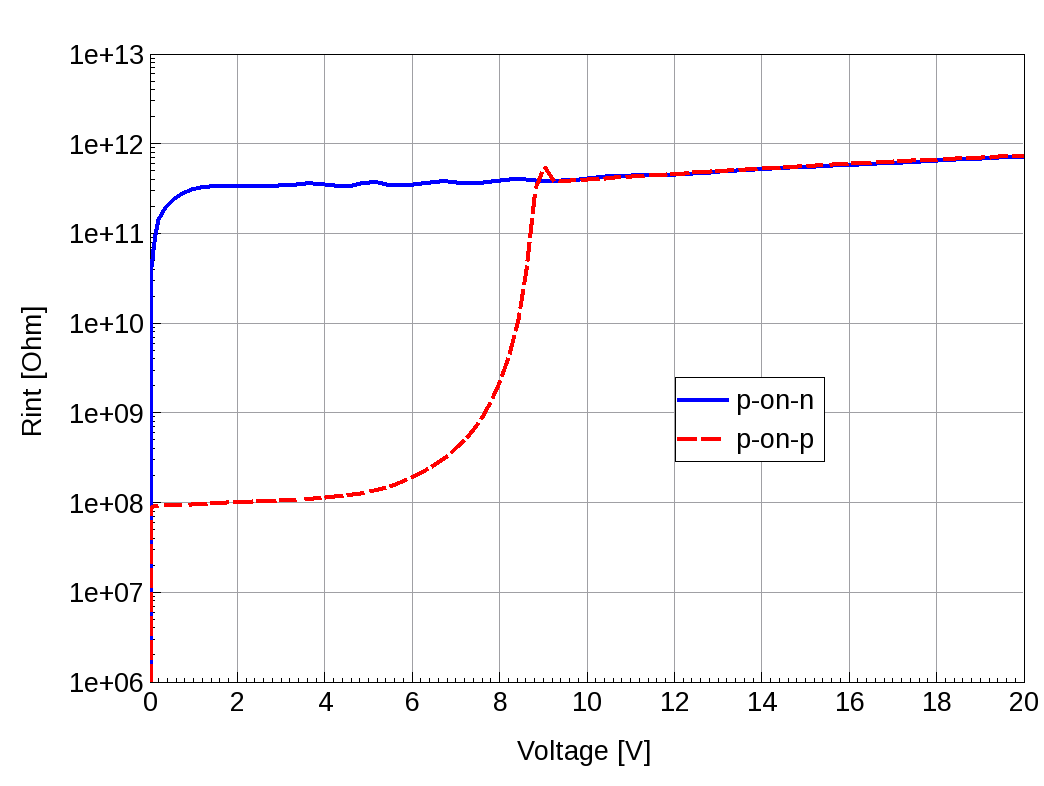}\label{Rint}}
     \caption{(a) Simulated $CV$ curves at $f$ = 10 kHz for the two sensor types and their full depletion voltages $V_{\textrm{fd}}$ determined from the crossing point of the linear fits to the dynamic and plateau regions of the curves. (b) Interpixel resistances $R_{\textrm{int}}$ as a function of bias voltage for the two sensors.}
     \label{fig:fig9a}
\end{figure}

\begin{figure}[tbp]
     \centering
     \subfloat[][$V$ = 5 V.]{\includegraphics[width=.65\textwidth]{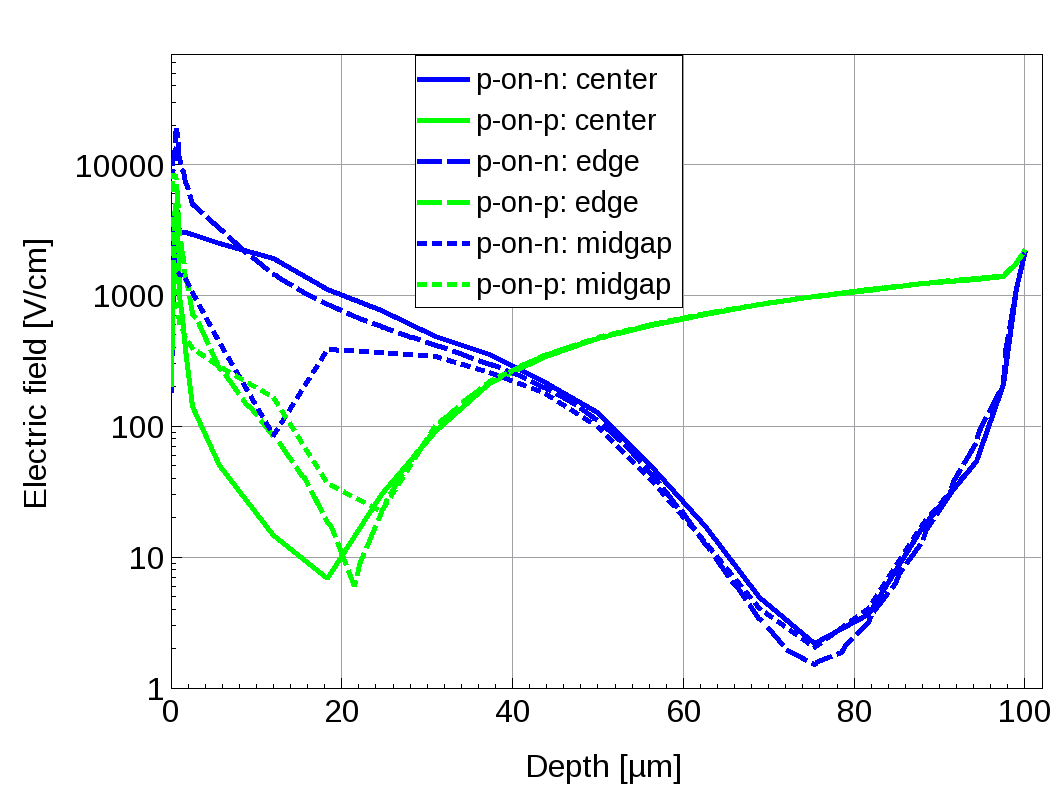}\label{E5V}}
     \par\vfill
     \subfloat[][$V$ = 90 V.]{\includegraphics[width=.65\textwidth]{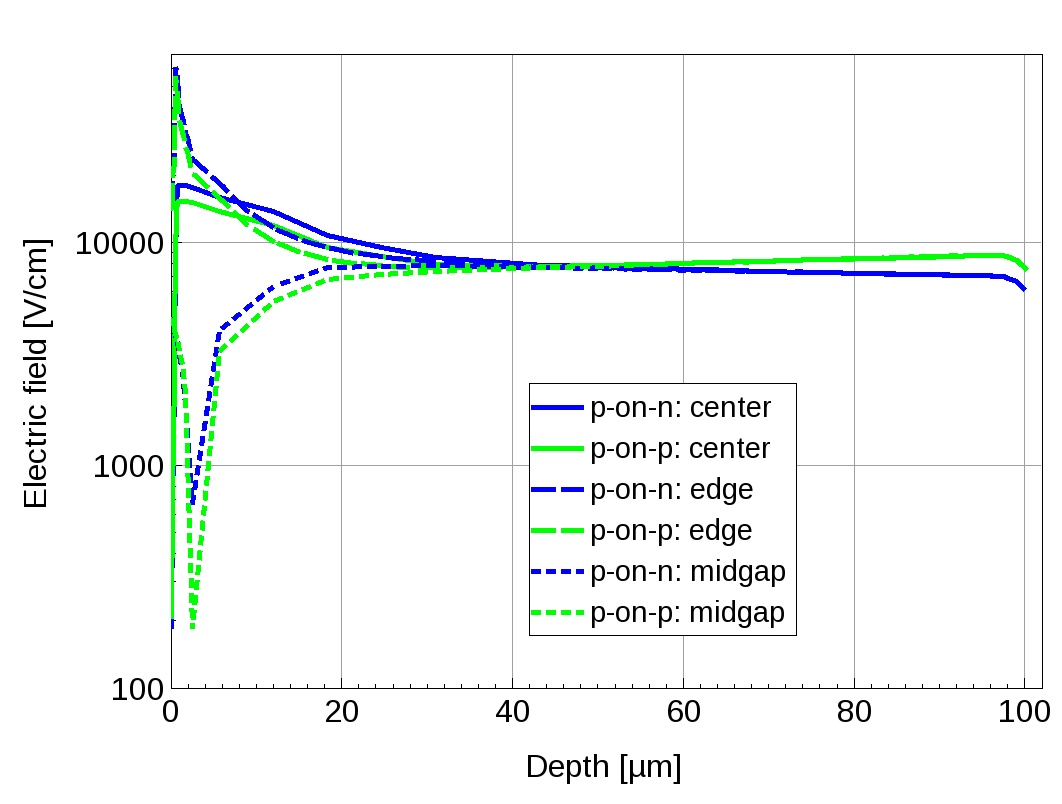}\label{E90V}}
     \caption{Simulated electric field distributions on logarithmic scale through the device bulk for the two sensor types at the (a) lower and (b) higher part of the investigated voltage range. The cuts were made in the middle of the centermost pixel (center), at the pixel implant edge (edge) and at the center of the inter-pixel gap (midgap). The ratios of the electric field maxima $E$(p-on-p)/$E$(p-on-n) at the pixel edge are $\sim$43$\%$ and $\sim$87$\%$ for 5 V and 90 V, respectively.}
     \label{fig:9c}
\end{figure}

\begin{figure}[tbp] 
\centering
\includegraphics[width=.6\textwidth]{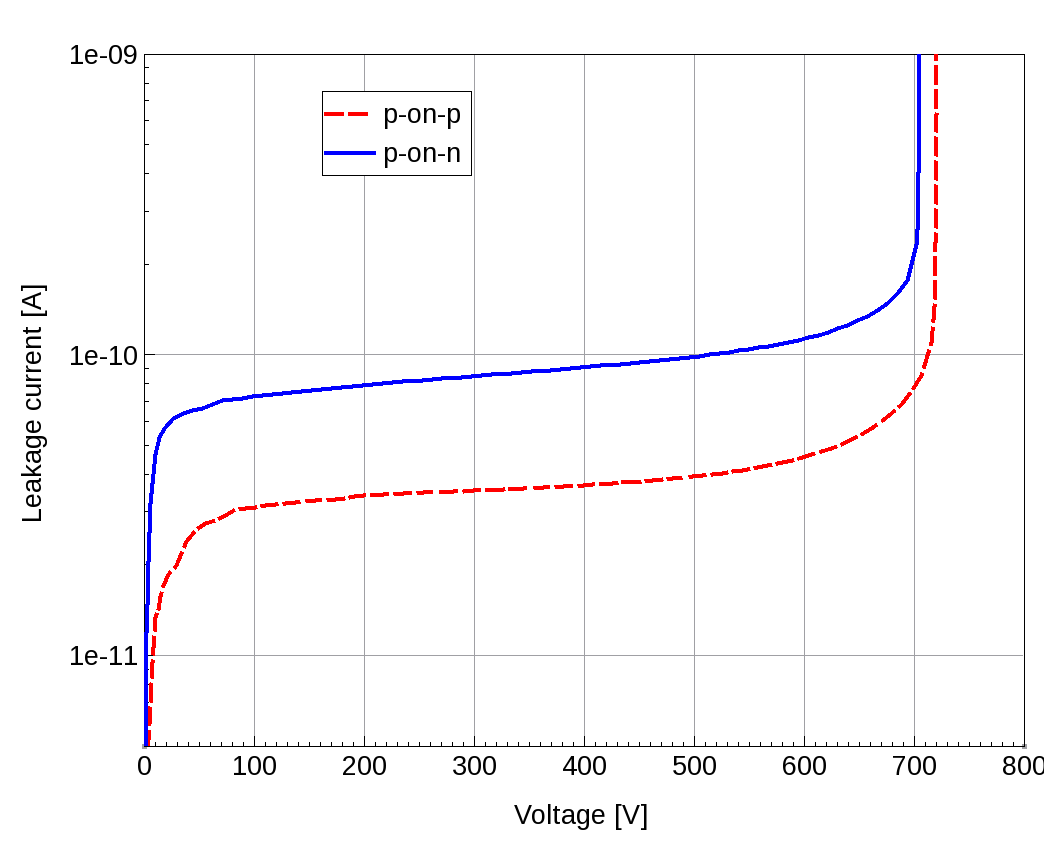}
\caption{Simulated $IV$ curves for the two sensor types and their breakdown voltages $V_{\textrm{bd}}$. The $V_{\textrm{bd}}$ is $\sim$16 V lower for the p-on-n sensor.} 
\label{fig:Vbd}
\end{figure}

Even though the bulk doping concentration of the p-on-p sensor was set to a considerably higher value than for the p-on-n, as shown in table~\ref{tab:2}, 
the capacitance-voltage ($CV$) simulations of the two sensors, presented in figure~\ref{CV}, resulted in a lower full depletion voltage $V_{\textrm{fd}}$ for the p-on-p, namely $V_{\textrm{fd}}$(p-on-p) $\approx$ 7 V and $V_{\textrm{fd}}$(p-on-n) $\approx$ 11 V. These were in line with the measured $CV$-results in reference \cite{Kalliop2013}. The $CV$ characteristics are different for the p-on-p and the p-on-n because for the p-on-p the depletion starts from the non-segmented side with immediate increase of the depletion depth at small voltages, while for p-on-n it starts from the segmented side, for which part of the voltage is used to deplete the area between the pixels, slowing the extension of the effective depletion depth and the change of capacitance, as presented in figures~\ref{CV} and~\ref{E5V}. Also seen in figure~\ref{E5V} is that the depletion process of the p-on-p sensor is further enhanced by the extension of the electric field from the electrode sides due to the potential differences produced by the high doping gradients at the pixel edges. This is reflected in figure~\ref{CV} where around 6 V the rate of the bulk depletion changes from equal to the p-on-n sensor to almost instantaneous. 
Figure~\ref{E90V} shows that 
at higher values of the reverse bias voltage the electric field maximum in the p-on-p sensor is at the pixel side, thus good resolution and low collection times can be expected. 
As can be seen from figure~\ref{fig:9c}, the peak electric fields at the pixels are higher in the p-on-n sensor, especially at lower voltages. This would lead to expect generally better breakdown behaviour for the p-on-p sensor. However, as presented in figure~\ref{fig:Vbd}, the rapid increase of the electric fields with voltage at the pixel edges in the p-on-p sensor leads eventually to a breakdown behaviour that is very close the p-on-n sensor. Since the measured leakage current was roughly 3-fold higher in the p-on-n sensor \cite{Kalliop2013}, this required a tuning of the carrier lifetimes (discussed in detail in the following section) to be reproduced in figure~\ref{fig:Vbd}. 

The interpixel resistance simulations in figure~\ref{Rint} display over 
three orders of magnitude higher resistance for the p-on-n sensor until full depletion is reached in the p-on-p sensor. With the $V_{\textrm{fd}}$ result from figure~\ref{CV} it can be seen from the $R_{\textrm{int}}$ curve of the p-on-p sensor that after the bulk is fully depleted it still takes $\sim$2 V to deplete the inter-pixel region. After $\sim$9 V the pixel isolation between the two sensors is identical. Thus, no significant differences in charge collection due to the electric field distribution and the inter-pixel isolation in the two sensor types can be expected after about 9 V of bias voltage.    
%
%
%
\subsubsection{Charge collection}
\label{Charge collection}
To reproduce the charge injection by the 800 keV proton energy presented in figures~\ref{fig:fig4} and \ref{fig:fig5a}, the range and the energy loss from the SRIM simulations were used as the input parameters for the TCAD simulation where the injection point was set to the center of the sensor's non-segmented backplane, opposite to the centermost pixel in figure~\ref{fig:8b}. Since the measured charge was recorded from a single pixel using a trigger that excluded all two pixel or larger cluster events, the chosen charge injection position for the simulation was considered to give sufficient approximation of the real measurement situation.  
The simulation temperature and the SiO$_2$/Si interface charge density $Q_{\textrm{f}}$ were set to typical values for a non-irradiated silicon detector, 293 K and $3\times10^{10}$ $\textrm{cm}^{-2}$, respectively. For the studied voltage range, from 15 V to 100 V, the role of carrier diffusion in fully depleted sensors was found to be negligible. Thus, for simplicity the charge collection simulations were carried out with diffusion switched off.

\begin{figure}[tbp] 
\centering
\includegraphics[width=.65\textwidth]{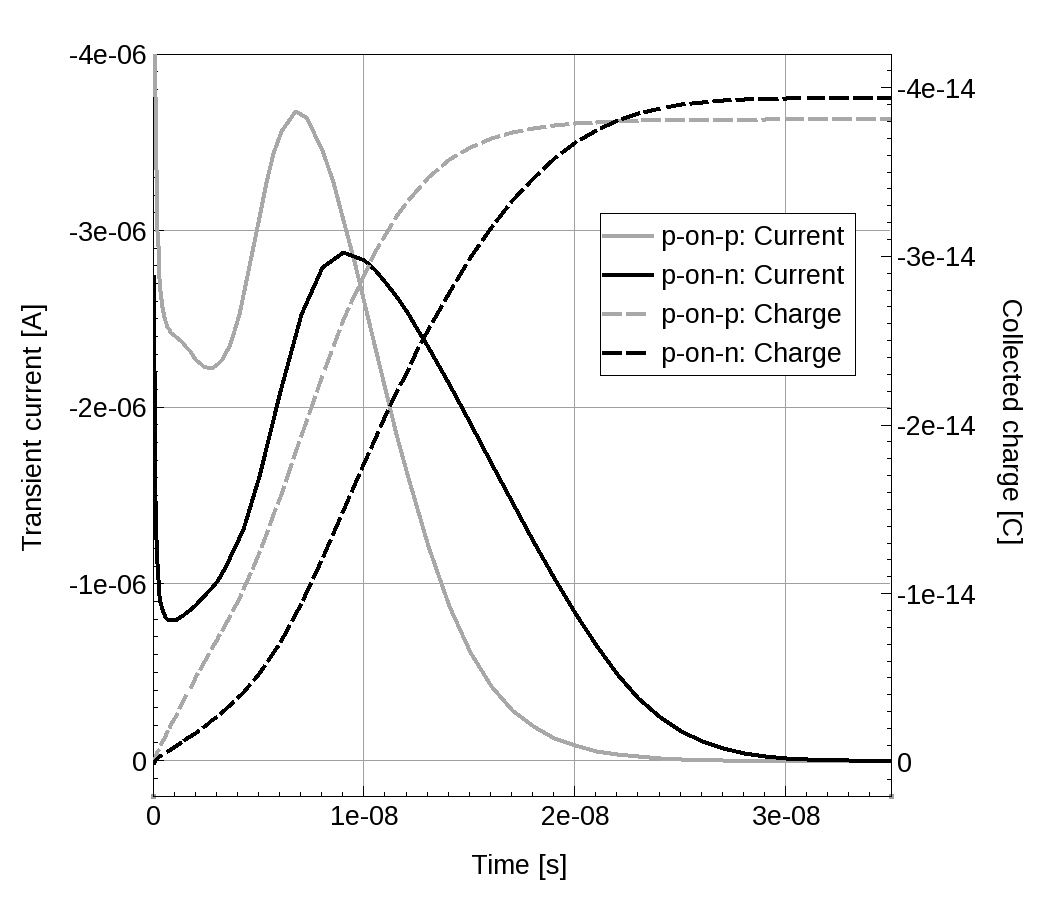}
\caption{Simulated transient currents and collected charges in the p-on-p and p-on-n sensors of figure~\ref{fig:8b} at $V$ = 23 V. Charge in the range of a 800 keV proton was injected from the middle of the backplane and the charges collected at the center-most pixel are plotted. The carrier lifetimes correspond to the tuned leakage current in figure~\ref{fig:Vbd}. By using elementary charge and the average energy of 3.62 eV required to create an $e$-$h$ pair in room temperature, the collected charge scales to energy as $4\times10^{-14}~\textrm{C}\sim900~\textrm{keV}$.} 
\label{fig:fig11}
\end{figure}

\begin{figure}[tbp] 
\centering
\includegraphics[width=.65\textwidth]{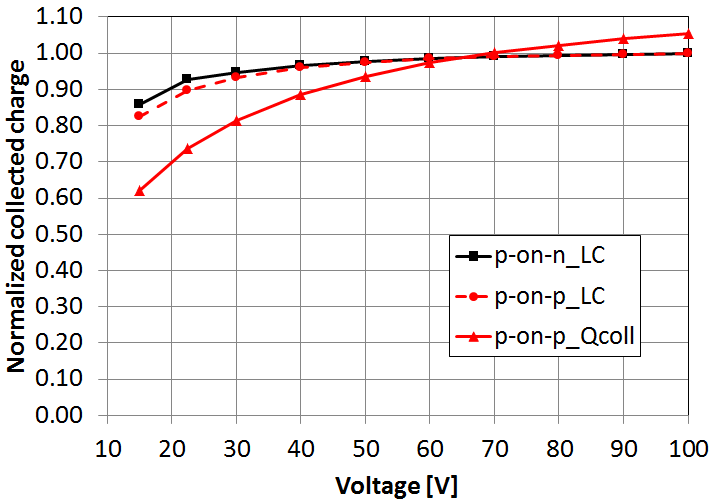}
\caption{Simulated charges collected at the centermost pixel of the structure in figure~\ref{fig:8b} for p-on-p and p-on-n sensors. Normalization is done to the charge collected by the p-on-n sensor at $V$ = 100 V. LC indicates the curves produced by the charge carriers with lifetimes tuned to reproduce experimentally matching leakage current ratio in the two sensors. $Q_\textrm{coll}$ corresponds to the carrier lifetime tuning to move the relative charge collection behaviour between the two sensors closer to the measurement in figure~\ref{fig:fig4}.}
\label{fig:fig12}
\end{figure}

The initial charge collection simulations with the charge carrier lifetimes tuned to reproduce the experimentally observed difference in the leakage current (presented in figure~\ref{fig:Vbd}) displayed similar qualitative behaviour with the curves in figure~\ref{fig:fig4}. As seen in figures~\ref{fig:fig11} and~\ref{fig:fig12} the p-on-n sensor is collecting higher charges at low voltages and at higher voltages the collected charges in the two sensors converge. However, whereas at e.g. 20 V the measured collected charge was $\sim$25\% higher in the p-on-n sensor the corresponding difference was only $\sim$5\% in the simulation. Also the initial collected charges for the both sensors around 15 V are considerably higher than in the measurement.  
Any effect from the charge sharing in the measurement was ruled out by the charge collection conditions described above. Also, the interpixel resistance simulations in section~\ref{Electrical characteristics} displayed an identical pixel isolation in the two sensor types after $\sim$10 V. Thus, difference in this regard between simulation and measurement could not be considered to explain the smaller difference in the simulated collected charges between the two sensor types. 
In addition, since the backplane metallization thicknesses and doping diffusion depths were essentially identical in the measured sensors, no contribution to the charge collection behaviour from the difference in the scattering of protons or in charge carrier losses at the non-active region could be expected. 

Thus, further study of the charge collection was focused on the charge carrier lifetimes in the two sensor types. The initial objective of the carrier lifetime tuning was to reproduce simultaneously lower leakage current and lower charge collection at low voltage 
for the p-on-p sensor. In the case of a hole dominated transient signal due to the backplane charge injection, this was addressed by increasing the electron lifetimes while decreasing the hole lifetimes with respect to the p-on-n sensor. The effect of the tuning for the carrier collection is seen in figure~\ref{fig:fig11} where the differences between the two sensor types in electron (small $t$) and hole (larger $t$) contribution to the current signal are clearly visible\footnote{It should be noted that since the linear energy transfer of the Bragg peak of the 800 keV proton proved problematic to fine-tune in the TCAD simulation and since only the relative charge collection between the two sensors was investigated, a deposited energy within $\sim$100 keV of the nominal value was used for the simulation study.}. 

By decreasing the lifetimes of both carrier types it is possible to shift significantly the charge collection evolution with voltage towards the measured values in the p-on-p sensor, as displayed by the solid red curve (p-on-p$_\textrm{Qcoll}$) in figure~\ref{fig:fig12}. The ratio of the collected charges by the 800 keV proton injection in the two sensors at 15 V now matches the measurement, while the collected charge relative to the maximum collection at the same voltage is within 10\% of the measured p-on-p sensor. The crossing point of the two curves is now only $\sim$10 V from the measured and at voltages beyond the crossing point the p-on-p collects somewhat higher charges than p-on-n, which is also seen in figure~\ref{fig:fig4}. Thus, this gives an indication that higher trapping rate of holes in the p-bulk at low voltages leads to the observed charge collection difference in the two sensors. At increased bias voltages, the total drift time of the carriers is reduced due to the higher electric field leading to reduced number of trapped carriers in the detector and higher collected charges \cite{Kramberger2002p}.

By applying this approach also to the p-on-n sensor would provide a means to tune the low voltage charge collection as well as the charge collection and leakage current ratios between the two sensors close to measured values. However, even though the mutual ratio of the leakage current in the sensors would be preserved, its absolute values would increase beyond measured values in the process. 
Thus, the tuning of the carrier lifetimes should be considered as an effective approach to reach an estimation of the higher hole trapping in the p-on-p sensor that could explain the measured charge collection behaviour. The implementation of specific trap levels (i.e. deep traps with high hole trapping probability and low contribution to the leakage current \cite{Radu15}) to the simulation with characteristical carrier capture cross-sections, activation energy and concentrations, as extracted from the measured data of their properties, would be more precise approach, but in lack of such data is left to a later study. 

\begin{figure}[tbp] 
\centering
\includegraphics[width=.65\textwidth]{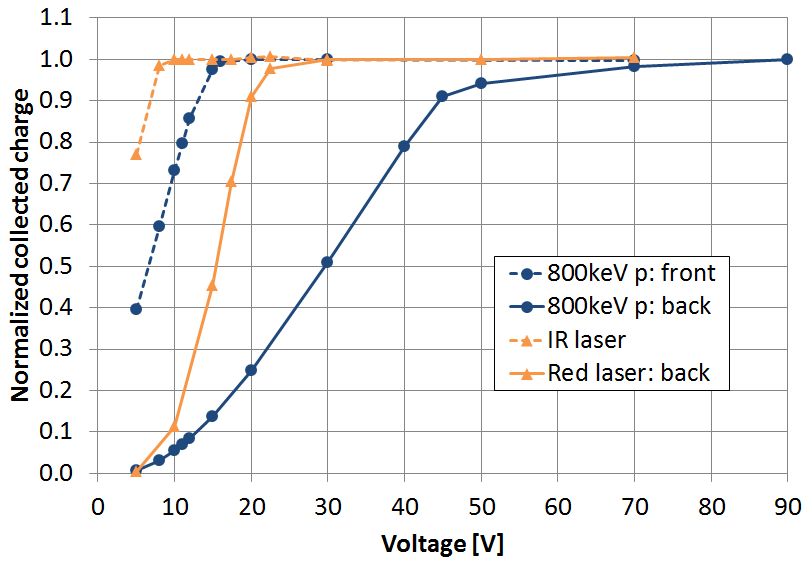}
\caption{Simulated charges collected at the centermost pixel of a 2-dim. 3-pixel p-on-n sensor similar to a diagonal slice of the 3-dim. structure in figure~\ref{fig:8b}. Normalization is done to the maximum charge collected at each injection. Charge injections from the front surface and the backplane are indicated accordingly. IR = infrared, p = proton.} 
\label{fig:fig13}
\end{figure}

Finally, a simulation study of the voltage required for a complete charge collection ($V_\textrm{cpl}$) is presented in figure~\ref{fig:fig13}. Since only general tendencies were investigated, a simplified 2-dim. p-on-n structure was applied for the charge injections from both the front surface and the backplane. Figure~\ref{fig:fig13} displays a strong dependence of $V_\textrm{cpl}$ on the charge carrier type and also a significant dependence on the size of the generated carrier cloud. Of the four studied charge injections, the IR laser signal (produced by both electrons and holes) is collected at $V_\textrm{cpl}$ = $V_\textrm{fd}$ $\approx$ 11 V. 
When a red laser (absorbed within 10 $\mu\textrm{m}$) of equal intensity is injected from the backplane (front surface injection resulted in a curve overlapping the IR laser curve) the signal produced by the hole drift more than doubles $V_\textrm{cpl}$. The 800 keV proton injection generates a considerably higher localized carrier density resulting in $V_\textrm{cpl}$ $\sim$15 V from the front surface injection (electron drift) and about 3-fold higher $V_\textrm{cpl}$ from the hole drift due to the backplane injection. Thus, regardless of the approximations of the 2-dim. structure\footnote{Collected charges in figure~\ref{fig:fig13} at low voltages are much lower than in the 3-dim. simulations due to the underestimation of the electric field evolution in the pixel sensor.} the results show that a transient signal produced by a hole drift (with $\sim$3-fold lower mobility to electrons) leads invariably to $V_\textrm{cpl}$ higher than $V_\textrm{fd}$. Additionally, a localized carrier density generated by an energy deposition in the range of hundreds of keV results in a decrease of the mobility of the carriers, due to the increase of carrier-carrier interactions, leading to a shift of $V_\textrm{cpl}$ to higher voltages. Hence, these two observations provide an interpretation of the measured $V_\textrm{cpl}$ behaviour in figure~\ref{fig:fig4}. 


\section{Conclusions}
\label{Conclusions}
Active edge silicon p-on-p pixel detectors of 100 $\mu\textrm{m}$ thickness were fabricated on 150 mm float zone silicon wafers at VTT. Spectroscopic characterization performed with a $^{241}$Am source and a mini X-ray tube showed that peaks of all edge pixels and center pixels are well aligned and the heights of the peaks are proportional to the effective volumes of the pixels. This indicates a good functionality of the edge pixels.

Investigation of the detector active volume as a function of pixel-to-edge distance revealed that at the applied test voltage of 20 V 
the sensitive region reaches its maximal possible size, given that the distance to the physical edge is less than 100 $\mu\textrm{m}$. 

For the TCAD simulation study of the p-on-p and p-on-n sensor types a 3-dimensional sensor structure was applied to model the circular-shaped pixels correctly. The electric field distributions in p-on-p sensor show that the electric field maximum is at the pixel edges, due to 
geometrical effect seen on all segmented devices, after full depletion is reached. Since the pn-junction in p-on-p device is at the non-segmented side, the electric field peaks at the pixels and the backplane lead to 
low full depletion voltage of the bulk. 
If the pn-junction is only at the pixel side, as in the p-on-n technology, this leads to higher $V_{\textrm{\small fd}}$ than in non-segmented devices
, due to the lateral expansion of the electric field that slows the extension towards backplane. The interpixel resistance simulations in the two sensor types displayed identical behaviour after full depletion. Also the difference in the simulated breakdown voltages of the two sensors was observed to be negligible. Thus, the location of the pn-junction at the non-segmented side in the p-on-p detector does not result in any significant advantage on this regard. 

The initial charge collection simulations from a backplane injection with carrier lifetimes tuned to reproduce experimentally matching leakage current ratio between the two sensors resulted in a qualitative agreement with measurements but with considerably smaller voltage dependence. The result displayed the prevalent effect of the more favourable weighting field in the p-on-n sensor over the higher electric field at the location of the charge injection in the p-on-p sensor at low voltages. The voltage dependent charge collection of the p-on-p sensor was possible to be adjusted closer to the measured behaviour by further decreasing the 
hole lifetimes. This effective approach showed that the lower charge collection at low voltages in the p-on-p sensor could be explained by the higher hole trapping in the p-bulk. Combined with the measured lower leakage current in the p-on-p sensor this would suggest the presence of deep level traps in the p-bulk with significant hole trapping probability and small contribution to the leakage current \cite{Radu15}. 
To maintain the agreement with the measured leakage current levels, further tuning of the simulated charge collection should then be carried out by including such traps.

%
The simulation of the observed effect of a larger $V_{\textrm{\small cpl}}$ with respect to $V_{\textrm{\small fd}}$, in both sensor types, revealed a strong dependence on the carrier type and on the degradation of the carrier mobility due to high localized carrier densities.  
%
%
%

To conclude, the measurement and the simulation study of the active edge p-on-p pixel detector displayed equal electrical and charge collection performance to the more typical p-on-n active edge pixel detector at bias voltages above $\sim$80 V.
\section*{Acknowledgements}
Access and operation of the Van de Graaff accelerator of the IEAP CTU Prague were supported by the MSMT Grant Research Infrastructure No. LM2011030 of the Ministry of Education, Youth and Sports of the Czech Republic.
%

\bibliography{mybibfile}

\end{document}